\input{aipcheck.tex}

\documentclass[final]{aipproc}

\layoutstyle{6x9}

\newcommand{\be}{\begin{equation}}
\newcommand{\ee}{\end{equation}}

\newcommand{\ba}{\begin{eqnarray}}
\newcommand{\ea}{\end{eqnarray}}

\def\lsim{\lower.5ex\hbox{$\; \buildrel < \over \sim \;$}}
\def\gsim{\lower.5ex\hbox{$\; \buildrel > \over \sim \;$}}

\newcommand\etal{\textit{et al.\ }}

\begin{document}

\title{The ``Supercritical Pile'' Model of GRB: Thresholds, Polarization, 
Time Lags }

\author{Demosthenes Kazanas}{
address={NASA/Goddard Space Flight Center, Greenbelt, MD 20771}
}

\author{Markos Georganopoulos}{
  address={NASA/Goddard Space Flight Center, Greenbelt, MD 20771}
}

\author{Apostolos Mastichiadis}{
  address={Dept. of Astronomy, University of Athens, Panepistimiopolis, 
Athens 15784, Greece}
%  ,altaddress={Dept. of Astronomy, University of Athens, Panepistimiopolis, 
%Athens 15784, Greece} % additional visiting address
}

\begin{abstract}
The essence of the ``Supercritical Pile''  model is a process 
%for converting the kinetic energy of GRBs into radiation; more specifically 
for converting the energy stored in the relativistic protons of a 
Relativistic Blast Wave (RBW) of Lorentz factor $\Gamma$ into electron 
-- positron pairs of similar Lorentz factor, while at the same time
emitting most of the GRB luminosity at an energy $E_p \simeq 1$ MeV. 
This is achieved by scattering 
the synchrotron radiation emitted by the RBW in an upstream located 
``mirror'' and then re-intercepting it by the RBW. The 
repeated scatterings of radiation between the RBW and the ``mirror'', 
along with the threshold of the pair production reaction $p \gamma 
\rightarrow p e^-e^+$, lead to a 
maximum in the GRB luminosity  at an energy $E_p \simeq 1$ MeV, 
{\sl independent of the value of $\Gamma$}. Furthermore, 
the same threshold implies that the prompt $\gamma-$ray emission is only 
possible for $\Gamma$ larger than a minimum value, thereby providing a
``natural'' account for the termination of this stage of the GRB as the RBW 
slows down. Within this model the $\gamma-$ray ($E \sim 100$ keV -- 1 MeV)
emission process is due to Inverse Compton scattering and it is thus 
expected to be highly polarized if viewed at angles $\theta \simeq 
1/\Gamma$ to the RBW's  direction of motion. 
Finally, the model also predicts lags in the light curves of the lower 
energy photons with respect to those of higher energy; these are  of purely 
kinematic origin and of magnitude $\Delta t \simeq 10^{-2}$ s, in agreement 
with observation. 
\end{abstract}

%\date{\today}

\maketitle

\section{Introduction}

The discovery of GRB afterglows by $BeppoSAX$ and the ensuing 
determination of their redshifts \cite{costa}, ushered 
a new era in GRB physics and in our understanding of their 
time development. While the issue of their distance and energetics
was settled by these observations, a number of issues concerning the 
physics of the RBWs that give rise to the GRB phenomenon still remain 
open as novel issues have been raised with the advent of observations
and accumulation of more GRB of known redshifts \cite{lamb}. Among the 
older issues that still remain open is that of the narrow distribution 
of the GRB peak energy $E_p$ \cite{malozzi}, in view of its sensitive  
dependence on the RBW Lorentz factor ($E_p \propto 
\Gamma^4$) within the synchrotron model of GRB. Another such 
issue is that of conversion of the energy stored  in relativistic
protons in the RBW into electrons. The presence of protons was 
necessary in the early RBW models \cite{rees}
for transporting and releasing the energy carried off by the 
GRB RBW to the distances demanded by observation ($R \sim 10^{16}$
cm). While more recent MHD models \cite{vlah, lyut} are immune from 
this  requirement (though they still need a mechanism for dissipating
the magnetic energy), sweeping and accumulating the ISM matter 
by these MHD flows will store a similar amount of energy into 
relativistic protons as do more conventional models, still demanding 
a mechanism for converting this energy to radiation. Models generally 
resolve
this issue by assuming the equipartition of the total energy density 
between protons and relativistic electrons of arbitrary distributions, 
as demanded by the need to account for their observed spectral 
characteristics. 

In our view, the most compelling argument in favor our model is that
it can answer in a rather straightforward way both the issue
of the  the limited range in $E_p$ and that of conversion
the energy stored into relativistic protons to radiation. 
At the same time, it has additional implications which seem to be in 
agreement with the mounting GRB phenomenology; these will be discussed
in the following sections.

\section{``Supercritical Pile'': The Thresholds}
%\subsection{The Kinematic and Dynamic Thresholds}

We assume the presence of a population of relativistic protons 
of form $n(\gamma_p) =n_0\,\gamma_p^{-\beta}$ on the frame 
co-moving with the RBW, i.e. moving with Lorentz Factor (LF) 
$\Gamma$ with respect to the observer (and also at zero angle to his/her 
line of sight; see \cite{kazan} for details).  
We consider synchrotron photons from $e^-e^+$ pairs
of energy $\gamma_e$, $\epsilon_s = b \, \gamma_e^2$ ($b$ is 
the value of the magnetic field in units of the
critical value $B_c \simeq 4 \cdot 10^{13}$ G; all energies 
measured in units of the electron mass $m_ec^2$) reflecting
off an upstream ``mirror'' and being re-intercepted by the RBW.
Their energy will now be $\epsilon_s^{\prime} = \Gamma^2 \, b \, 
\gamma_e^2$. The threshold for $e^-e^+$ pair production of these
photons with a proton of Lorentz factor $\gamma_p \simeq \gamma_e$
$b \, \Gamma^2 \, \gamma_e^3 = 2$. Assuming that the 
proton and the electrons are drawn from the relativistic thermal
post-shock particle population, $\gamma_e \simeq \gamma_p \Gamma$, 
leading to the {\it kinematic threshold} condition
\begin{equation}
b \, \Gamma^5_{\rm th} \simeq 2 ~~~~~ {\rm or} ~~~~~ \Gamma_{\rm th} 
\gsim \left( \frac{2}{b} \right)^{1/5} 
\label{kinematic}
\end{equation}
Assuming equipartition for the magnetic field for a GRB with total
energy $E = 10^{52} E_{52}$ erg, restricted to an angle $\theta \simeq
1/\Gamma$ leads to $\Gamma_{\rm th} \gsim 90 (E_{52}/R_{16})^{-1/12}$ 
%a value in agreement with that used in most models
. 

For this reaction network to be self-sustained, at least one
of the reflected synchrotron photons must pair produce with the 
protons on the RBW (after its reflection by the ``mirror'') 
to replace the electron that 
produced it. Therefore the plasma optical depth $\tau$ to the pair
producing reaction $p \gamma \rightarrow p e^-e^+$ must be
at least as large as the inverse of the number of synchrotron
photons produced by a given electron. An electron of energy 
$\gamma$ produces ${\cal N}_{\gamma} 
\simeq \gamma /b \gamma^2 = 1/ b \gamma$ photons, yielding
$\tau = n_{\rm com} \sigma_{p \gamma} \Delta_{\rm com} \gsim 
1/{\cal N}_{\gamma}$, where $ n_{\rm com}, \Delta_{\rm com}$
are the comoving density and width of the RBW. Considering
that the column density is a Lorentz invariant $n_{\rm com} 
\Delta_{\rm com} = n R$  and taking into account the kinematic
threshold relation (Eq. \ref{kinematic}) the {\it dynamic threshold} 
condition reads
\begin{equation}
 \sigma_{p\gamma} \,\Delta_{\rm com} \,  n_{\rm com} = 
\sigma_{p\gamma} \, R \,  n \gsim b \Gamma ~~~{\rm or} ~~~ 
\sigma_{p\gamma} \, R \, n \, \Gamma^4 \gsim 2~~.
\label{dynamic}
\end{equation}
This latter condition (and the physics behind it) are akin 
to those of those of a ``supercritical'' nuclear pile,
hence the nomenclature of this model. 
For the typical values of $n$ and $R$ used in association with 
GRBs, i.e. $n = 1 \; n_0$ cm$^{-3}$ and $R = 10^{16}\; R_{16}$ 
cm and considering that $\sigma_{p\gamma} \simeq 5 \, 10^{-27}$ 
cm$^2$, the criticality condition yields $\Gamma \gsim 375 \, 
(n_0 \, R_{16})^{-1/4}$
%, values consistent with the accepted parameter range  
(Eq. \ref{dynamic} is slightly different
from that given in \cite{kazan}; we would like to thank P. M\'esz\'aros 
for this correction which does not affect the results otherwise).

Assuming the width of the reflecting ``mirror'' to be thinner 
than the width of the RBW, blast waves with column densities 
higher than that implied by Eq. (\ref{dynamic}) will release 
the energy stored in relativistic protons explosively on times 
scales comparable to the RBW light crossing time scale; 
otherwise, the duration of the burst will be comparable to the 
time it takes the RBW to cross the width of the ``mirror''. In 
this case, prominent emission is halted until the proton column
has been built up significantly to conform to the dynamic 
threshold. For RBW with $\Gamma$ between those of Eqs. 
(\ref{dynamic}) and (\ref{kinematic}) and assuming the 
presence of a ``mirror'' $\gamma-$ray emission continues as long
as its LF is greater than that implied by the kinematic 
threshold (Eq. \ref{kinematic}). Eventually, when the value of
the LF drops below this value, $\gamma-$ray emission stops and
the RBW enters the stage of afterglow.

\section{The GRB Spectra}

Consider a RBW of Lorentz factor $\Gamma$. Because of the 
relativistic focusing of emitted radiation, we need 
only consider a section of the blast wave of opening half 
angle $\theta = 1/\Gamma$. The shocked electrons of the 
ambient medium (and pairs from the $p \, \gamma \rightarrow 
e^+e^-$ process) produce, as discussed above, synchrotron 
photons of energy $\epsilon_s \simeq b \, \Gamma^2$. 
These, upon their scattering 
by the ``mirror'' and re-interception by the RBW, are 
boosted to energy $\epsilon = \epsilon_s \, \Gamma^2 = b \,
\Gamma^4$ (in the RBW frame). These photons will then be 
scattered by the following electron populations: (a) By electrons 
of $\gamma \simeq 1$, originally 
contained in the RBW and/or cooled since the explosion. 
(b) By the hot ($\gamma \simeq \Gamma$), recently shocked 
electrons to produce inverse Compton (IC) radiation at energies 
correspondingly $\epsilon_1 \simeq b \, \Gamma^4$ and 
$\epsilon_2 \simeq b \, \Gamma^6$ at the RBW frame. 
(the SSC process will also yield photons at $\epsilon_{ssc} 
\simeq b \, \Gamma^4$, however it turns out that this 
is not as important and it is ignored here). 
At the lab frame, the energies of these three components, 
i.e. $\epsilon_s, \, \epsilon_1, \, \epsilon_2$ will be 
higher by roughly a factor $\Gamma$, i.e. they will be 
respectively at energies $b \, \Gamma^3,~b \, \Gamma^5$ 
and $b \, \Gamma^7$. Assuming that the process operates 
near its kinematic threshold, $b\, \Gamma^5 \simeq 2$, 
at the lab frame these components will occur at energies 
$\epsilon_s \simeq \Gamma^{-2}$, $\epsilon_1 \simeq 2 
\simeq 1$ MeV and $\epsilon_2 \simeq \Gamma^2 \simeq$ 
10 GeV $(\Gamma/100)^2$. This model therefore, produces 
``naturally'' a component in the $\nu F_{\nu}$ spectral 
distribution which peaks in the correct energy range. 
It also predicts the existence of two additional components at 
an energies $m_ec^2 \, \Gamma^2$ and $m_ec^2/\Gamma^2$. The
high energy emission has been observed from several GRBs \cite{dingus}. 

\section{Other Issues}

The simplicity by which this model deals with several
outstanding issues of GRB suggest that one should attempt
to test its viability by addressing additional GRB 
systematics and properties.

%\subsubsection{Time Lags}
A. {\it Time Lags}. It has been observed that, in general, the light
curves of soft photons lag with respect to those of the 
harder ones. A systematic study of these lags in a sample of GRBs with 
known redshift has determined that these lags range between $0.01 - 0.1$
sec, with their magnitude in inverse correlation to the
GRB peak luminosity \cite{norris}. The model we presented above can 
produce lags of the  order of magnitude observed: According to the 
basic premise of our model the emission at $E \sim 50 - 500$ keV 
is due to bulk IC scattering of synchrotron photons, by the ``cold'' 
electrons of the RBW, after been reflected  the ``mirror''. The 
eventual energy of these photons depends on the angle of their 
direction after scattering at the ``mirror'' with respect to the 
velocity vector of the RBW (this being highest for a ``head-on'' 
collision). Because the distance between the ``mirror'' and the 
RBW is of order $R/\Gamma^2$, one can easily estimate that the path
difference between the soft and hard photons are of the same order of 
magnitude i.e. $\Delta L \simeq R/\Gamma^2$. Therefore the corresponding 
time lags (assuming the observer to be along the direction of the velocity
of the RBW) should be $\Delta t \simeq \Delta L/c \Gamma^2 \simeq 
R/c \Gamma^4 \simeq 10^{-2.5} R_{16}/\Gamma_2^4$ sec, in agreement 
with observations. 

B. {\it Polarization}. The recent results of high ($80 \pm 20 
\%$) polarization of GRB 021211 \cite{boggs} has raised the 
interest of the community in this particular aspect of GRBs.
While models employing synchrotron radiation can produce 
at best polarization $\lsim 70\%$ (for totally uniform field 
geometry), models producing the 100 - 1000 keV radiation 
by the inverse Compton process can potentially produce 
polarization approaching $ 100 \%$ for particular orientation
of the observer \cite{dar, ghisel, eich03}. This is a purely 
geometric effect: Thomson scattering of unpolarized radiation 
to an angle $\theta = 90^\circ$ with respect to the incident
direction erases all electric field orientations but that 
perpendicular to the plane defined by the incident and 
scattering directions leading to $100 \%$ polarization in this
direction. Since in the lab frame this direction corresponds to 
an angle $\theta = 1/\Gamma$, the polarization should rise from
$0 - 100 \%$ in going from $\theta = 0$ to $\theta = 1/\Gamma$ 
and drop again for larger angles. However, in our model we
scatter not unpolarized radiation but the synchrotron radiation
from the RBW, which is itself polarized. This leads to a 
non-zero polarization even for angles close to $\theta = 0^\circ$
thus enhancing the probability that we observe a high polarization
signal.

C. {\it General Considerations} We would like to  point
out that the arguments presented above have completely ignored the
possibility of particle acceleration at the RBW of GRBs (a fundamental 
requirement of most models). We have dealt with the conversion in
pairs of  the energy stored only in the thermal population of protons. A
non-thermal component will ease the thresholds of Eqs. 
(\ref{kinematic}, \ref{dynamic}) and allow high energy emission  long
after the end of the prompt GRB phase, as it appears to be the 
case in some GRBs \cite{dingus}.

\end{document}